\documentclass{PoS}

\title{IceCube search for high-energy neutrinos produced in the precursor stages of gamma-ray bursts}

\ShortTitle{IceCube search for GRB precursor neutrinos}

\author{
  The IceCube Collaboration\footnote{For collaboration list, see PoS(ICRC2019) 1177.}\\
  {\itshape \href{http://icecube.wisc.edu/collaboration/authors/icrc19_icecube}{http://icecube.wisc.edu/collaboration/authors/icrc19\_icecube}}\\
  E-mail: \email{paul.coppin@icecube.wisc.edu}
}

\abstract{

While gamma-ray bursts (GRBs) were once believed to be a dominant source of astrophysical neutrinos, the lack of any significant correlation between high-energy photons and neutrinos has resulted in stringent limits on their neutrino fluxes. Previous IceCube searches for neutrinos from GRBs have generally focused on the prompt phase of GRBs. However, the higher density during the GRB precursor stage could lead to more hadronic interactions, and therefore be the predominant time of neutrino emission. We present results from an analysis of Fermi-GBM data to identify and characterize individual precursor flashes. Together with an up-to-date catalog of GRBs, we have made our results available via an online tool. An IceCube analysis method to search for neutrinos arriving coincident with GRB precursors has been developed. In this presentation, we compare the sensitivity of this analysis to previous IceCube analyses and discuss the implications in case a significant signal is found.\\

\vspace{4mm}
{\bfseries Corresponding authors:}
\speaker{Paul Coppin}$^{1}$, Nick van Eijndhoven$^{1}$\\
{$^{1}$ \itshape Vrije Universiteit Brussel}

}

\FullConference{36th International Cosmic Ray Conference -ICRC2019-\\
		July 24th - August 1st, 2019\\
		Madison, WI, U.S.A.}

\usepackage{amsmath}

\makeatletter
\renewcommand\@makefntext[1]{\leftskip=3em\hskip-1em\@makefnmark#1}
\makeatother

\begin{document}

\section{Introduction}\label{sec:intro}
Gamma-ray bursts (GRBs) are the most energetic explosions of electromagnetic radiation observed in our Universe. After their initial discovery, it was soon revealed that the duration of GRBs follows a bimodal distribution \cite{Zhang}. Further research confirmed that GRBs can have at least two types of progenitors. Outbursts lasting longer than two seconds are thought to occur when the core of a very massive star undergoes gravitational collapse. This collapsar model has been confirmed for a number of long  duration GRBs by the detection of the accompanying broad-lined type Ic supernovae \cite{TypeIcSN}. Short GRBs, lasting less than two seconds, are thought to originate when two co-orbiting compact objects, where at least one is a neutron star, merge after spiraling towards each other. Compelling evidence for this model was only recently obtained when the LIGO and Virgo collaborations reported the detection of gravitational waves from a neutron star merger in coincidence with a short GRB observed by the Fermi-GBM and INTEGRAL telescopes \cite{BNS_merger}.\par
Despite the existence of two distinct progenitor classes, the physics of the outburst can be described within one model \cite{Zhang}. For both progenitor classes, a highly relativistic fireball will be launched due to accretion of matter onto the central engine, and multiple shells of ejecta can be successively expelled with varying Lorentz factors. The collisions that subsequently take place when fast shells catch up with slower ones will lead to the acceleration of particles. Accelerated electrons will emit synchrotron radiation, the features of which are consistent with the observed gamma-ray light curves \cite{Zhang}. Other particles, such as protons, will likewise be accelerated in these collisions. Based on the Hillas criterion \cite{Hillas}, GRBs should be capable of accelerating particles up to the highest energies $\left(10^{20}\ \mathrm{eV}\right)$.\par
If GRBs are sources of ultra-high-energy cosmic rays, the interaction of high-energy protons near the source would result in the production of neutrinos carrying $\sim$4\% of the primary proton energy. The IceCube collaboration has performed a number of searches aiming to identify these GRB neutrinos. In a search covering 1172 GRBs, no statistically significant signal was found \cite{IC_GRB}. With the lack of any significant correlation, limits were placed on both the neutrino flux as well as the model parameters of GRBs.\par
It should however be noted that the neutrino flux upper limit only applies to the prompt phase of GRBs. In roughly 15\% of all GRBs, a small emission episode is observed a few seconds to a few hundred seconds before the start of the main outburst \cite{SwiftPrecursors, BATSEPrecursors}. These precursor flashes indicate that central engine activity is, in some cases, already ongoing before the start of the prompt emission. Various models have been proposed for different progenitors. In the case of long GRBs, one prevalent model predicts that precursors occur when an initial weak jet is unable to punch through the circumburst medium \cite{JetFallBackModel}. As this mechanism implies a temporarily chocked jet\footnote{Gamma-ray emission is in this case still expected if the jet stalls close below the photosphere \cite{ChockedJet}.}, the largest neutrino fluxes are in this case expected not during the prompt, but during the precursor phase.\par
For this reason, we aim to perform an IceCube search looking for neutrinos from GRB precursors. An all new GRB catalog has been constructed for this purpose and is discussed in section \ref{sec:grbweb}. Individual precursor flashes were identified by analyzing data from the Fermi-GBM detector as outlined in section \ref{sec:gbm}. In section \ref{sec:ice}, we then present the sensitivity of our proposed IceCube analysis.

\section{The new GRB catalog}\label{sec:grbweb} 
To perform an IceCube GRB analysis, we start by composing an all inclusive catalog of gamma-ray bursts. Currently, the two prominent GRB detectors are the Fermi Gamma-ray Burst Monitor (GBM) and Swift Burst Alert Telescope (BAT). There are however numerous other telescopes, such as INTEGRAL and KONUS-Wind, contributing to the detection of GRBs. While gamma-ray detectors are essential to identify bursts and capture the characteristics of the light curves, the uncertainty on the position of the bursts is typically several degrees \cite{GBM_detector, SwiftDetector}. If the burst was observed by multiple satellites, then a better localization can be obtained by triangulating their signals, as is done by the InterPlanetary Network (IPN). An alternative approach is to observe the afterglow of the GRB. This is achieved by a wide array of telescopes, together covering the electromagnetic spectrum from x-ray to radio. In addition to providing an improved localization, these observatories can also provide complementary information, such as the redshift of the object.\par
Almost all observatories report their observations via GCN-circulars to facilitate follow-up observations. Unfortunately, GCN-circulars are text messages, often written by hand, that are not designed to be machine readable. In order to still use the information from this extensive resource, we developed a parsing tool that uses regular expressions to automatically extract the relevant information from the circulars. All entries are then saved in a MySQL database. This database is extended by adding catalogs of specific detectors which are directly available online, such as the Fermi (GBM \& LAT), Swift (BAT, XRT \& UVOT) and IPN catalogs. For consistency, all catalogs are converted and stored in a uniform format. A single summary table of GRBs is then composed, grouping all listings of a single GRB across all tables. Only the preferred value of each variable, for instance that with the lowest relative uncertainty, is saved to the final summary table.\par
In spirit with the original version of GRBweb \cite{OldGRBweb}, the full database is publicly available online at \href{https://icecube.wisc.edu/~grbweb_public}{https://icecube.wisc.edu/~grbweb\_public}. The website contains not only data about GRBs, but also Python and text-based extraction tools for downloading the data. Additionally, metadata are available detailing the origins and formats of the GRB data. Users are encouraged to send feedback if they have questions, comments or if they want to request new features.\par
At the moment, this new database contains over six thousand GRBs, about 500 of which have a redshift measurement. These numbers are constantly increasing as the full database gets updated on a weekly basis. Apart from certified GRBs, subthreshold bursts are also listed in the database and new features are continuously being added as the project progresses.

\section{Identifying precursors in Fermi-GBM data}\label{sec:gbm}
\begin{figure}
\includegraphics[width=\textwidth]{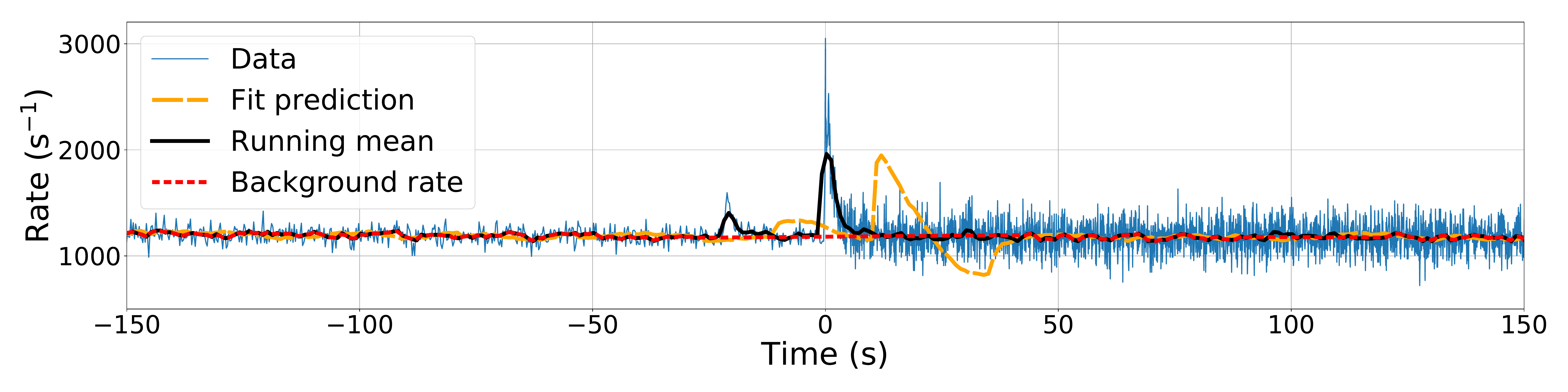}
\includegraphics[width=0.985\textwidth]{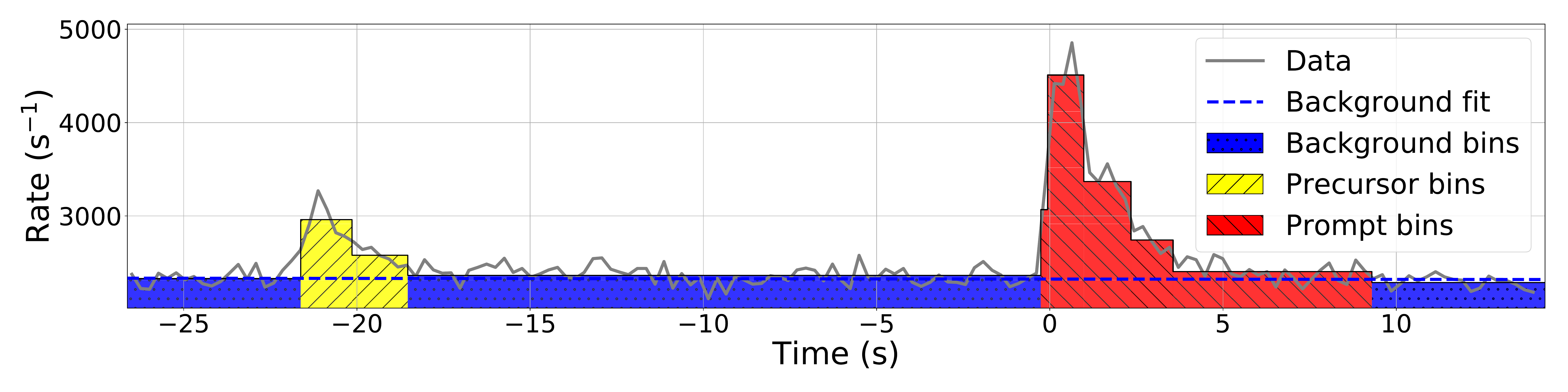}
\caption{Top: Characterization of the GBM background rate for GRB trigger bn120308588 and detector n8. The running mean of the rate (black) is compared to the prediction from a fit to prior data points (orange). Regions in which the two distributions match are used to fit the background rate (red). Bottom: Bayesian block histogram of the same GRB (bn120308588), using the combined data from detector n4 and n8. Bins belonging to the precursor and prompt phase are respectively marked in yellow and red.}
\label{fig:light_curve}
\end{figure}

Having constructed a catalog of GRBs, we still require a listing of the precursors of those bursts. In the past, various searches have been performed to identify and characterize precursor emission episodes \cite{SwiftPrecursors, BATSEPrecursors}. Unfortunately, no up-to-date catalog of GRB precursors is currently available. The most recent search by the Fermi-GBM collaboration \cite{Sylvias_thesis} examined bursts up until the end of 2013, though there are no current plans to update this analysis to include more recent bursts. Given that the IceCube detector was only fully completed in 2011, this would severely restrict the data that could be used in our proposed IceCube analysis. We therefore performed an analysis on the publicly available data\footnote{Accessible via \href{https://heasarc.gsfc.nasa.gov/FTP/fermi/data/gbm/bursts}{https://heasarc.gsfc.nasa.gov/FTP/fermi/data/gbm/bursts}.} from the Fermi-GBM detector to identify and characterize precursors episodes in all bursts since 2011. Apart from having data publicly available, the GBM detector was chosen as it observes more GRBs per year ($\sim$240) than any other detector. In the remainder of this section, we outline the methods we used and give a short overview of our results.\par
The Fermi-GBM telescope consists of twelve relatively low energy (8 keV to 1 MeV) thalium doped sodium iodide (NaI) detectors and two high-energy (200 keV to 40 MeV) bismuth germanate (BGO) detectors. These are pointed in varying directions, allowing the GBM to monitor the full region of the sky that is unocculted by the Earth. GBM data of GRBs are available as a set of files listing the photon counts in each of the detectors as a function of time. The highest temporal resolution can be achieved using the TTE (time tagged events) files, which contain the arrival time of individual photons with a 2 $\mu$s precision. Due to bandwidth restrictions, TTE data are only available from $\sim$50 seconds before to 200 seconds after the burst\footnote{The exact time range varies on a burst-per-burst basis.}. CTIME (continuous high time resolution) data, containing a binned light curve with a nominal binsize of 0.256 seconds, is used to extend this observation window to 500 seconds before and after the burst \cite{GBM_detector}.\par
An example of a typical GRB light curve is shown in the top of Fig. \ref{fig:light_curve}. $t=0$ s marks the time at which the GBM detector was triggered by the burst. A distinct emission episode, visible as a bump in the mean rate (black line), can however be observed about twenty seconds before the trigger time. To distinguish these types of probably precursor episodes from random background fluctuations, we started by characterizing the background rate. On average, the background rate of each detector is about 1000 Hz, though strong fluctuations are possible depending on the orientation of the detector. If the rate is stable, i.e. when there is no burst, the rate should be well approximated by a linear curve on time scales shorter than 50 s. We therefore performed a linear fit to the rate in the interval $\left[t-35\ \mathrm{s},\ t-10\ \mathrm{s}\right]$ and used that fit to predict the rate at time $t$. This procedure was performed for all times $t$ that are an integer in seconds, leading to the orange curve. We set the estimated background rate equal to the running mean of the rate (black curve) in regions where the running mean effectively matches the predicted value. Intermediate regions are covered by linearly interpolating between the nearest estimates of the background rate (red line).\par
Once the background rate is known, the Bayesian block \cite{BayesianBlocks} light curve is constructed for all individual detectors which were triggered by the burst. In addition, we construct a single Bayesian block light curve using the combined data from these detectors\footnote{If the trigger involved more than three detectors, only those three detectors with the smallest angular separation with respect to the burst are used.}. This results in a light curve histogram with variable bin-width in which both the number of bins and their widths are optimized \cite{BayesianBlocks}. An example of such a light curve is given at the bottom part of Fig. \ref{fig:light_curve}. Bins containing a significant excess are then identified by comparing their true rate to the rate of their background parameterization. If the average rate in a bin exceeds the background rate by 30 Hz, it is tagged as a signal bin if this excess is also observed coincidently in the individual light curves of at least two of the NaI detectors. This approach is consistent with that employed in the previous precursor search by the GBM collaboration \cite{Sylvias_thesis} and strongly restricts the probability that a background fluctuation is misidentified as a precursor event.\par
The background subtracted light curve is then composed, counting only the events that are contained in signal bins. As in Fig. \ref{fig:light_curve}, the possibility exists that there are multiple emission episodes, i.e. periods of increased gamma-ray emission separated by a quiescent period\footnote{We require a quiescent period of at least two seconds to separate two emission periods.} in which the rate returns to the background level. If this is the case, we identify the prompt emission as the emission episode that contains the largest fluence, corresponding to the background subtracted integral of the red blocks in Fig. \ref{fig:light_curve}. Any preceding emission episode is then tagged as a precursor if its fluence is less than a third\footnote{A similar cut was imposed in all previous precursor searches \cite{SwiftPrecursors, BATSEPrecursors, Sylvias_thesis}.} of the fluence of the prompt emission.\par 
By applying this method to the GRBs detected by Fermi-GBM between the start of 2011 and the end of 2018, we obtained a set of 172 precursor events. These precursors are distributed over 145 GRBs\footnote{GRBs were allowed to have more than one precursor.} out of a sample of 1843 bursts, implying that about 8\% of all detected GRBs are observed to have at least one precursor episode. Roughly half of the precursors we identified were also seen in the previous Fermi-GBM search \cite{Sylvias_thesis} when comparing overlapping years. Readers interested in obtaining the selected sample can do so by accessing the database mentioned above. The emission times of all precursor episodes are contained in the \textit{Fermi\_GBM\_precursors} table, together with the start time of the prompt emission episode.

\section{IceCube search for precursor neutrinos}\label{sec:ice}
The obtained selection of precursors can now be used to identify time windows of opportunity to search for coincident neutrino events. Two types of analyses are envisioned to characterize possible correlations. Given the time windows at which precursors were observed for each individual GRB, one method is to only consider those neutrinos that arrive during one of the precursor episodes. This would result in a very low background search, as we know exactly when and where to look for coincident neutrinos. An alternative approach would be a (stacked) search for neutrinos in the time interval $\left[t_p-\Delta t, t_p\right]$, where $t_p$ marks the start of the prompt emission. The size of the examined universal time interval $\Delta t$ could in this case be based on the generic properties of our precursor sample. While the GRB specific temporal information of the precursors is disregarded in this search, this also means that it is applicable to all GRBs and thus not only to those observed by the Fermi-GBM detector.\par
IceCube is currently the world's only neutrino telescope which has been capable of distinguishing an astrophysical flux of neutrinos \cite{AstroNu}. Therefore, it is the prime instrument for our proposed GRB precursor neutrino search. The IceCube detector \cite{IC_detector} is located at the geographic South Pole and was completed in 2011. It consists of 5160 detection modules, which are distributed over 86 vertical strings and embedded between depths of 1450 and 2450 meter in the polar ice sheet. Neutrinos interacting in the ice or the nearby bedrock are detected by means of the Cherenkov radiation emitted by the resulting relativistic charged particles. Based on the signature of the particles in the detector, the energy, direction and flavor of the initial neutrino can be reconstructed \cite{IC_detector}.\par
Every second about 2700 atmospheric muons, produced in cosmic-ray air showers, trigger the IceCube detector. As this background would overwhelm the expected neutrino signal, a stringent filter is applied, selecting only events that have a high likelihood of originating from a neutrino interaction. This leads to a reduced rate of 6.7 mHz. The resulting data sample contains mostly atmospheric neutrinos, as neutrinos produced in astrophysical sources cannot on an event-by-event basis be distinguished from those produced in cosmic-ray air showers\footnote{Unless the atmospheric neutrino is accompanied by muons created in the same air shower.}.\par
To quantify correlations between IceCube neutrinos and GRB precursors, we use an unbinned maximum-likelihood method. The likelihood we employ is given by
\begin{equation}
 \mathcal{L}=\frac{\left(\hat{n}_s+n_b\right)^N}{N!}\ e^{-\hat{n}_s-n_b}\prod_{i=1}^{N}{\frac{\hat{n}_sS(x_i)+n_bB(x_i)}{\hat{n}_s+n_b}}\ ,
 \label{eq:Likelihood}
\end{equation}
where $n_b$ is the number of expected background events, $N$ is the number of observed events and $\hat{n}_s$ is the number of signal events that will be fit. $S(x_i)$ and $B(x_i)$ are respectively the signal and background probability distributions (PDFs) evaluated for the $i$-th neutrino event. Both $S$ and $B$ consist of spatial, energy and temporal terms \cite{IC_GRB}. A flat time profile is used for the signal PDF, which drops to zero outside the examined time window. Eq. \eqref{eq:Likelihood} can be minimized for $\hat{n}_s$ for each GRB separately, or the data from all GRBs can be combined in a stacking analysis, resulting in a single $\hat{n}_s$ value for the total data sample.

\begin{figure}[!tbp]
  \centering
  \begin{minipage}[t]{0.49\textwidth}
    \includegraphics[width=\textwidth]{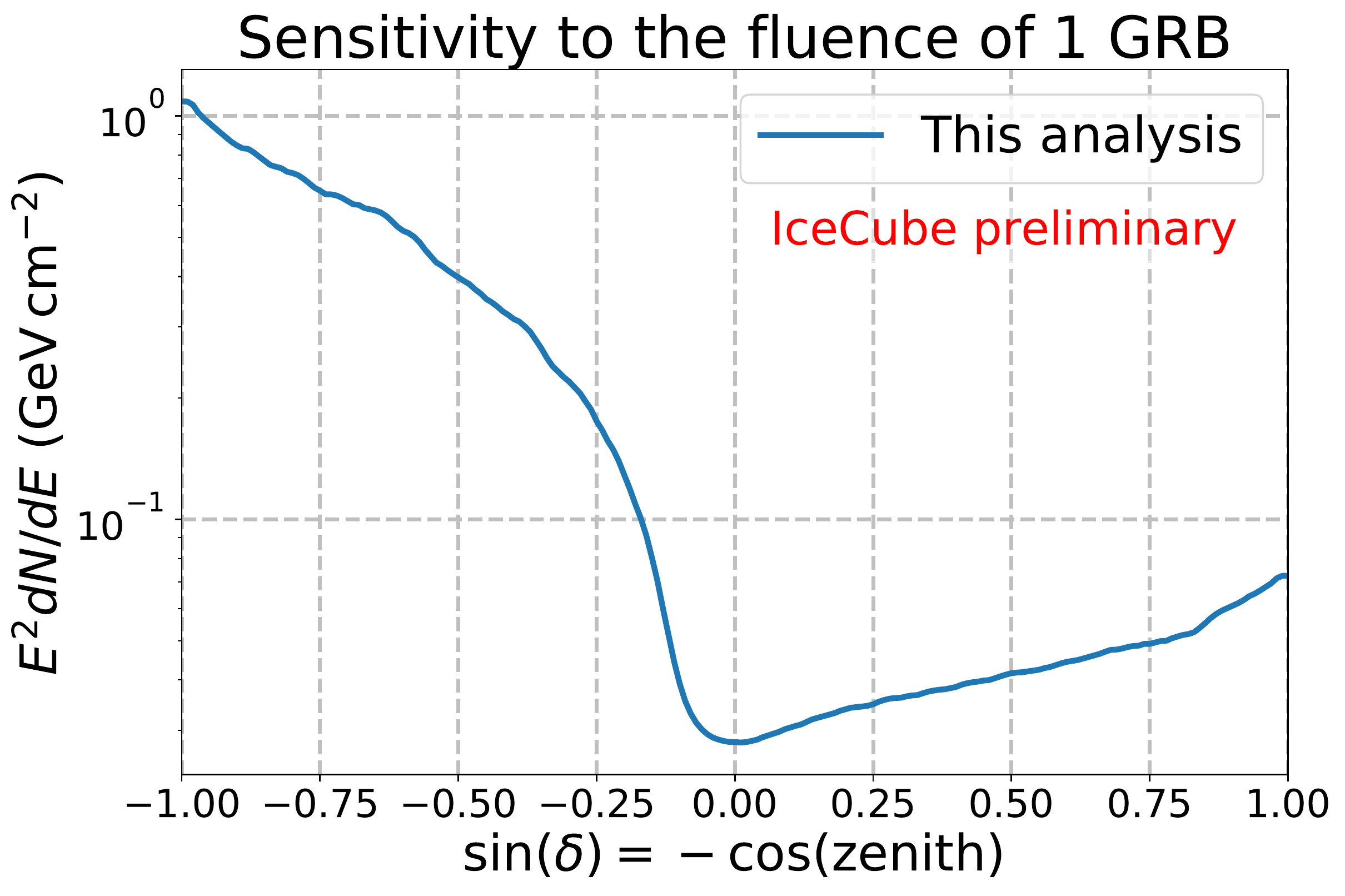}
    \caption{Sensitivity of our analysis to the neutrino fluence of a single GRB as a function of its declination and valid for any time window $\Delta t<1000$ s.}
    \label{fig:Sens_dec}
  \end{minipage}
  \hfill
  \begin{minipage}[t]{0.49\textwidth}
    \includegraphics[width=\textwidth]{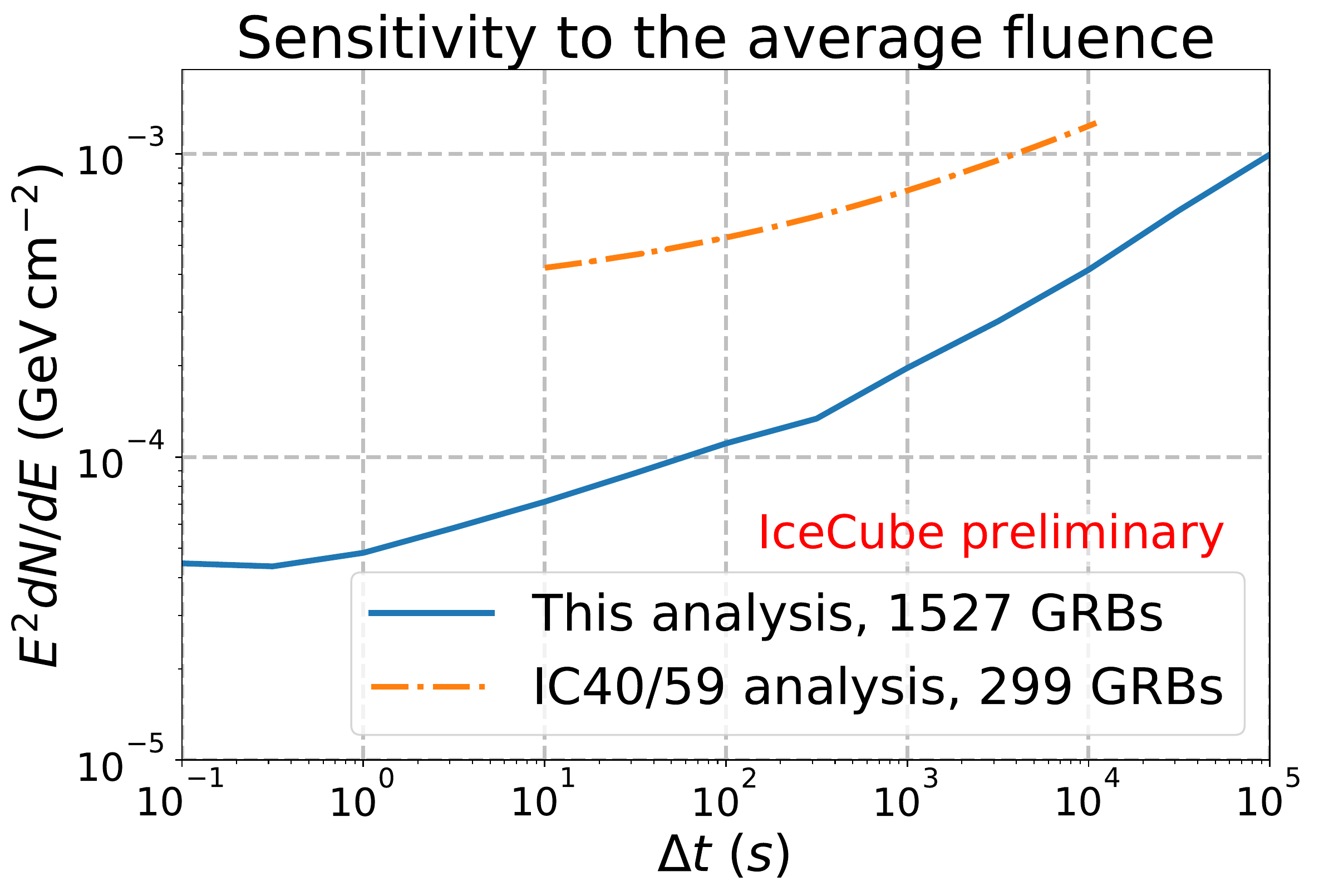}
    \caption{Comparison between our analysis and the previous IceCube analysis \cite{IC4059GRBPaper}, showing the sensitivity to the average neutrino fluence from respectively 299 and 1527 GRBs.}
    \label{fig:Sens_stack_dt}
    \end{minipage}
\end{figure}

The main factor determining the sensitivity of our analysis to a single GRB is the declination $\delta$ of the burst. When looking at the Northern hemisphere, IceCube is shielded from atmospheric muons by the Earth. A looser event selection can thus be used than in the Southern hemisphere, increasing the probability that a GRB neutrino passes the filter used to remove background events. This effect can be seen in Fig. \ref{fig:Sens_dec}, where $\sin(\delta)>0$ correspond to the Northern hemisphere. Another important effect is the Earth absorption of neutrinos, which becomes non-negligible for neutrinos with energies exceeding 100 TeV. A trade-off thus occurs between muon shielding and neutrino absorption, causing the optimal sensitivity to occur around the horizon where $\sin(\delta)=0$.\par
Fig. \ref{fig:Sens_dec} pertains to any search in which the examined time window $\Delta t$ is less than 1000 s, since in this region of parameter space, the number of background events $n_{b}$ arriving in temporal and spatial coincidence is $\ll 1$. As a result, the average number of signal events $\lambda$ required to obtain a best fit $\hat{n}_s$ value larger than zero in 90\% of all trials is independent of $\Delta t$ and corresponds to $\lambda\approx2.3$ events. The same fluence is thus required irrespective of $\Delta t$, as long as $\Delta t<1000$ s.\par
This reasoning does however not apply to a stacking analysis, as the total number of background events is no longer negligible when combining hundreds of bursts. In this case, the size of the considered time window $\Delta t$ does influence the sensitivity, whereas there no longer is a declination dependence as the sample of GRBs will be distributed uniformly across the sky. The most stringent limits on the neutrino fluence from GRB precursors were set by IceCube in 2012 \cite{IC4059GRBPaper}, using data from the partially completed detector with 40 and 59 strings denoted as IC40/59. In Fig. \ref{fig:Sens_stack_dt}, we compare the sensitivity of our analysis (blue solid line) to that of the IC40/59 analysis (orange dashed line). Both lines correspond to the average neutrino fluence each GRB would need to emit if we assume that all GRBs contribute equally\footnote{Alternatively, a redshift or photon fluence based weighting scheme could be used.} to the total neutrino fluence. Our analysis is shown to be more sensitive for all considered time window sizes $\Delta t$. The improvement is in large part due to the use of the full detector geometry, an improved event selection and a roughly 5 times larger GRB sample. Further improvements are still expected as the error estimates on GRB localizations from o.a. the GBM have for now been estimated in a conservative way.

\section{Conclusion and outlook}\label{sec:conl}
Various models suggest that the primary time of neutrino emission does not coincide with the prompt phase of gamma-ray bursts, but occurs during the earlier precursor stage of the bursts. We therefore intend to perform an IceCube analysis to look for neutrinos arriving in temporal and spatial coincidence with GRB precursors. To enable our analysis, we have composed an up-to-date catalog of GRBs, which combines the data from a wide range of observatories. This catalog is publicly available online at \href{https://icecube.wisc.edu/~grbweb_public}{https://icecube.wisc.edu/~grbweb\_public}.\par
As a second step, we have analyzed the publicly available Fermi-GBM data from all bursts between the start of 2011 and the end of 2018. For each burst, we parameterized the background rate and composed a Bayesian block light curve to identify precursor activity. Out of a sample of 1843 GRBs, we found that 8\% have at least one precursor episode. An update to this analysis is envisioned to further characterize the properties of the obtained precursor sample. In addition, we plan to perform further analyses of the light curves to identify other possible precursors using looser selection criteria, similar to the approach of the previous Fermi-GBM search \cite{Sylvias_thesis}.\par
Finally, we presented our foreseen IceCube analysis method and examined its sensitivity. For a single burst, the main variable driving the sensitivity is the declination of the bursts. When the examined time window $\Delta t<1000$ s, the fluence upper limits which can be obtained are independent of $\Delta t$ as the search is essentially background free. We then compared the sensitivity of our analysis to that of the previous IceCube search, combining data from 1527 and 299 GRBs respectively. While our analysis could already provide a significant improvement on the current results, we anticipate further improvements to the sensitivity via the use of less conservative error estimates on the localization of bursts.

\bibliographystyle{ICRC}
\bibliography{references}

%

\end{document}